\documentclass[prl,amsmath,amssymb,twocolumn]{revtex4}

\usepackage{graphicx}
\usepackage{subfigure}
\usepackage{adjustbox}
\usepackage{bm}
\usepackage{color}
\usepackage{braket}
\usepackage{standalone}
\usepackage{multirow}
\usepackage{tikz}
\usepackage{mathrsfs}
\usepackage{dsfont}
\usepackage[colorlinks,bookmarks=true,citecolor=blue,linkcolor=red,urlcolor=blue]{hyperref}

\newcommand{\bea}{\begin{eqnarray}}
\newcommand{\eea}{\end{eqnarray}}

\begin{document}

\title{Non-Hermitian Topological Sensors}

\author{Jan Carl Budich$^1$}
\email{jan.budich@tu-dresden.de}
\author{Emil J. Bergholtz$^2$}
\email{emil.bergholtz@fysik.su.se}
\affiliation{$^1$Institute of Theoretical Physics${\rm ,}$ Technische Universit\"{a}t Dresden and W\"{u}rzburg-Dresden Cluster of Excellence ct.qmat${\rm ,}$ 01062 Dresden${\rm ,}$ Germany\\\
$^2$Department of Physics, Stockholm University, AlbaNova University Center, 106 91 Stockholm, Sweden}
\date{\today}

\begin{abstract}
We introduce and study a novel class of sensors whose sensitivity grows exponentially with the size of the device. Remarkably, this drastic enhancement does not rely on any fine-tuning, but is found to be a stable phenomenon immune to local perturbations. Specifically, the physical mechanism behind this striking phenomenon is intimately connected to the anomalous sensitivity to boundary conditions observed in non-Hermitian topological systems. We outline concrete platforms for the practical implementation of these non-Hermitian topological sensors (NTOS) ranging from classical meta-materials to synthetic quantum-materials.
 
\end{abstract}

\maketitle

{\it Introduction.---} High-precision sensors represent a key technology that is of ubiquitous importance in both science and everyday life. In the quest for future sensors, a main challenge is to identify viable scenarios, where basic physical observables become highly susceptible to changes in the quantity to be detected, yet in a controllable fashion. Prominent examples along these lines include microcavity sensors \cite{cavitysensor,cavitysensor2,cavitysensor3,cavitysensor4,singleatomionsensor}, metal oxide semiconductor field-effect transistor (MOSFET) based sensors \cite{MOSFET}, mechanical transducers \cite{mechtrand}, graphene sensors \cite{graphenesensor}, superconducting quantum interference devices (SQUIDS) \cite{Squids} and magnetometers \cite{magnetometers}.

Dissipative systems effectively described by non-Hermitian (NH) Hamiltonians can exhibit a high susceptibility to small perturbations of their complex energy spectrum that has no counterpart in closed Hermitian settings \cite{BerryDeg,Heiss,EPreview,BBC,lee,Xi2018,yaowang,Henning}. As a first step towards using the unique algebraic properties of NH matrices \cite{BerryDeg,Heiss,EPreview} for sensing \cite{EPsensorTheory}, an enhancement of precision in sensors operating at NH spectral degeneracies known as exceptional points (EPs) has been claimed \cite{EPsensorExp,EPsensorExp2}. Furthermore, in NH lattice systems with many degrees of freedom, a striking spectral sensitivity has recently been found in the context of NH topological systems, where phase transitions driven by small changes in the boundary conditions have been predicted and observed \cite{BBC,lee,Xi2018,yaowang,review,Ghatak,ElectricBBC,ElectricBBC2,QdynBBC,SzameitScience2020}.

\begin{figure}[t]
    \centering
    \includegraphics[width=\linewidth]{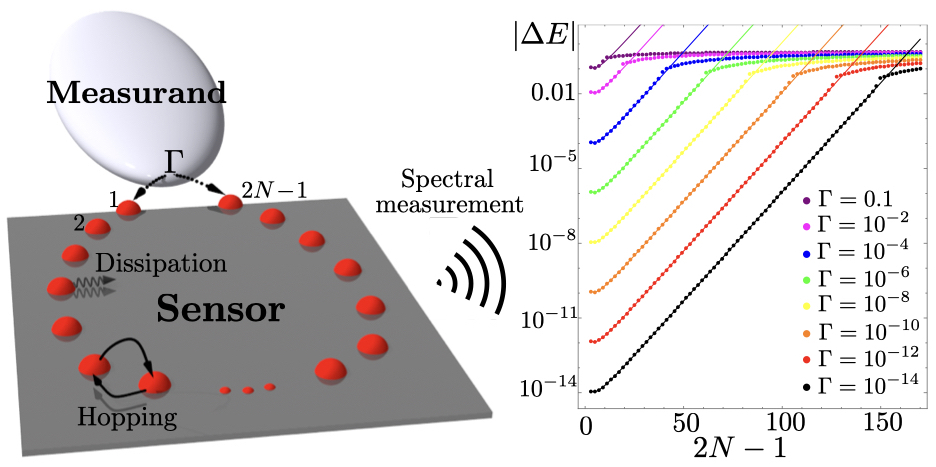}
    \caption{Left: Illustration of the non-Hermitian topological sensor (NTOS) setup consisting of one-dimensional chain with $2N-1$ sites in open ring geometry, described by an effective NH Hamiltonian. The measurand interferes with the coupling $\Gamma$ between the ends of the open ring (sites $1$ and $2N-1$). The simplest measurement signal is the energy shift $\Delta E$ of the lowest eigen-mode. Right: Exponential scaling of $\lvert \Delta E\rvert$ with system size $N$ of the NTOS based on the NH topological insulator model Eq.~(\ref{eqn:nhssh}) for different values of $\Gamma$. Dots are exact numerical energies which are in striking agreement with the analytical perturbative expression Eq.~(\ref{eqn:deltaE}) (solid lines). Parameters are $t_1=t_2=2\gamma=1$. In a wide parameter regime, the quotient of $\lvert\Delta E/\Gamma\rvert$ gives a good estimate for the sensitivity $\mathcal S$ of the NTOS.}
    \label{fig1}
\end{figure}

Here, drawing intuition from NH topological phases \cite{review,gong,NHarc,Budich2019, Yoshida2019, EPringExp,Kawabata2019}, we identify and study a widely applicable mechanism for drastically enhancing the sensitivity of a novel type of devices coined non-Hermitian topological sensors (NTOS). These systems are designed such that the physical quantity to be detected (measurand) effectively couples to the boundary conditions of an extended system of $2N-1$ lattice sites \cite{footOdd}, e.g. by modifying the coupling $\Gamma$ between the ends of the NTOS with a tunneling barrier (see Fig.~\ref{fig1}). Quite remarkably, in this scenario the energy $E_0$ of a characteristic bound state is then found to exhibit the exponential sensitivity
\begin{align}
\mathcal S=\frac{\partial E_0}{\partial \Gamma} =  \kappa\, \exp({\alpha N}),
\label{eqn:one}
\end{align}
where $\alpha, \kappa$ are model-specific constants. Below, we demonstrate both analytically and numerically that the exponential amplification of the sensitivity corresponding to $\rm{Re}[\alpha]>0$ is a generic and robust phenomenon that does not rely on any fine-tuning or symmetry and is as such immune to local perturbations including random disorder in the NTOS. Furthermore, we show how Eq.~(\ref{eqn:one}) manifests in the response signal of an NTOS device, thus elevating the exponential sensitivity to an enhanced precision in noise-limited experiments.

To illustrate the broad range of physical platforms in which NTOS may be realized from existing building blocks, we outline several implementations in the classical as well as in the quantum realm, including a detailed analysis of as system of coupled optical ring-resonators, as well as a discussion of topological electric circuits \cite{Albert2015,ElectricBBC,ElectricBBC2}, a quantum condensed matter setting of coupled quantum dots \cite{loss,Crooker2002,Kagan2016}, and a topological cold atom systems with dissipation \cite{Bloch2008,Goldman2016,Cooper2019}.

{\it Exponential amplification mechanism.---} We now derive and explain with theory the exponential scaling in system size parameter $N$ (cf. Eq.~(\ref{eqn:one})), i.e. in the number of physical degrees of freedom of the NTOS. To this end, we discuss a generic class of NH topological insulator models that exhibit the following basic phenomenology, noting that a specific case study will be presented further below. We consider systems that for an odd number of sites $2N-1$ exhibit a localized boundary state with energy $E_0(\Gamma)$, satisfying $E_0(0)=0$, i.e. an exact zero-energy edge mode in the limiting case of open boundary conditions \cite{BBC, SOM}. The corresponding unnormalized right- and left-eigenvectors are then of the general exponential form expressed in terms of the position (site) $j=1, \ldots ,2N-1$ in the lattice with non-vanishing components  
\begin{equation}
\braket{j|\Psi_{R,0}}=(r_R)^{\frac{j+1}{2}}\ ,\  \ \braket{\Psi_{L,0}|j}= (r_L)^{\frac{j+1}{2}},\ \ j \textrm{ odd} \label{endstates},
\end{equation}
where $\mu = R, L$ denote right- and left-eigenstates localized around $j=1$ if $\lvert r_\mu \rvert < 1$ and around $j=2N-1$ if $\lvert r_\mu \rvert >1$, and where the amplitude of the boundary mode vanishes for even values of $j$ \cite{SOM}. It is worth emphasizing that, in stark contrast to the Hermitian realm where $\lvert \Psi_{L,0}\rangle = \lvert \Psi_{R,0}\rangle$, right- and left-eigenvectors {\emph{generically differ}} in NH systems. In fact, the extreme scenario of an overlap 
\begin{equation}
\mathcal O_{LR} = \langle\Psi_{L,0}\vert \Psi_{R,0}\rangle/(\|\Psi_{L,0}\| \| \Psi_{R,0}\|)
\label{eqn:overlap}
\end{equation}
 that is exponentially small in system size  ($N$) represents a crucial ingredient for the exponential amplification at the heart of the proposed NTOS. An intuitive picture for this phenomenon is that in the considered NH scenario, a {\emph{single}} zero-energy state may completely change its spatial location when switching from a right- to a left-eigenstate description. 

To see this more quantitatively and derive Eq.~(\ref{eqn:one}), we now consider a coupling between the first ($j=1$) and last ($j=2N-1$) sites with a matrix element $\Gamma$, i.e.  $\Delta H=\Gamma(\ket{1}\bra{2N\!-1}+\ket{2N\!-1}\bra{1})$. A straight forward calculation to leading order perturbation theory in $\Gamma$ yields \cite{SOM}

\begin{align}
\Delta E &\! \approx\! \frac{ \bra{\Psi_{L,0}}\Delta H\ket{\Psi_{R,0}}}{\braket{\Psi_{L,0}|\Psi_{R,0}}} = \nonumber\\
&\Gamma \frac{(r_L r_R\! - \!1)(r_L^{N-1}\!\!+\!r_R^{N-1})}{(r_Lr_R)^N - 1}
\,~\overset{N \gg 1}{\longrightarrow}~\, \Gamma\, \kappa\,  e^{\alpha N} ,
\label{eqn:deltaE}
\end{align}
where the combination of the unperturbed right- and left-eigenstates entering the {\emph{biorthogonal}} expectation values occurring in Eq.~(\ref{eqn:deltaE}) is crucial, thus representing a natural generalization of perturbation theory to the non-Hermitian regime. The final expression in Eq.~(\ref{eqn:deltaE}) describes the leading asymptotic behavior of $\Delta E$ in the limit of large system size that is readily derived by comparing the competing powers of $r_L$ and $r_R$ \cite{SOM}. While $|\kappa(r_L,r_R)| > 0$, the real part of  the crucial parameter $\alpha(r_L, r_R)$ changes its sign corresponding to different physical regimes: As long as both the left- and right-eigenstates are localized at the same end, i.e. ${\rm sign}(\log(|r_R|))={\rm sign}(\log(|r_L|))$ (cf. Eq.~(\ref{endstates})), ${\rm Re}[\alpha]$ is negative and $\Delta E$ decays exponentially with system size, while $\mathcal O_{LR}$ remains of order $1$. This is the behavior intuitively expected and familiar from Hermitian systems. However, as soon as the left- and right-eigenstates are localized at opposite boundaries, i.e. ${\rm sign}(\log(|r_R|))\neq{\rm sign}(\log(|r_L|))$, ${\rm Re}[\alpha]$ becomes positive, driven by an exponentially small $\mathcal O_{LR}$ (see Eqs.~(\ref{eqn:overlap}-\ref{eqn:deltaE})), thus hallmarking a response that grows exponentially with system size (cf. Eq.~(\ref{eqn:one})).

{\it Case study of microscopic model.---}
To illustrate this phenomenology with a concrete microscopic model, and to quantify the validity regime of the perturbative result (\ref{eqn:deltaE}), we consider a NH Su-Schrieffer-Heeger (SSH) chain \cite{ssh1979, Lieu2018} with alternating nearest neighbor hopping amplitudes $t_1\pm\gamma$ and $t_2$, respectively. However, we stress that our main findings are of relevance beyond this specific setting \cite{footModel}. In reciprocal space, the model with periodic boundaries is described by the Bloch Hamiltonian
\begin{align}
H_S(k) =  \left(t_1 + t_2 \, \textrm{cos} k, \, t_2 \, \textrm{sin}k + i \gamma, \, 0 \right) \cdot \boldsymbol \sigma,
\label{eqn:nhssh}
\end{align}
where length is measured in units of the lattice spacing, and $\boldsymbol \sigma$ is the vector of Pauli matrices  acting on a sublattice degree of freedom with sublattices consisting of the sites with odd (even) index $j$. For an odd total number of sites $2N-1$ and open boundary conditions, there is an exact $E=0$ state of the form (\ref{endstates}) with $r_R=-(t_1-\gamma)/t_2$ and $r_L=-(t_1+\gamma)/t_2$ \cite{BBC,SOM}. To the open chain, we again add the coupling $\Delta H$ between the end sites.
 
In Fig.~\ref{fig1}, we plot the absolute value of the energy shift $\Delta E = E_0(\Gamma) - E_0(0)$ due to the presence of a finite coupling $\Gamma$ of the mode with smallest absolute value of energy for various values of $N$ and $\Gamma$. We have chosen the parameters $t_1=t_2=2\gamma=1$, noting that all qualitative results are robust against changing these parameters (see discussion below). The exact numerical energies (dots) are found to be essentially identical to the approximate analytical result (\ref{eqn:deltaE}) for a wide range of $N$ and $\Gamma$, yielding an exponentially amplified signal over many orders of magnitude. This quantitatively corroborates to striking accuracy our above general analysis. 

The perturbative energy shift in Eq.~(\ref{eqn:deltaE}) is real-valued for the NH SSH model, and the numerically obtained exact energies share this property to a remarkable precision until the energy shift reaches the scale set by the bulk gap. This is reflected in Fig.~\ref{fig1}, where deviations form the perturbative results (solid lines) occur at roughly similar $\Delta E$ independent of the coupling strength $\Gamma$. 

Generally, we find an exponential enhancement of the sensitivity $\mathcal S$ whenever $||t_1|-|t_2||<|\gamma |<||t_1|+|t_2||$. Interestingly, this desirable regime precisely corresponds to the parameter range with a non-trivial spectral winding number \cite{gong}
\begin{align}
\nu = \frac{1}{2\pi i} \oint_{-\pi}^{\pi}\text{d}k  \, \frac{\partial}{\partial_k} \log\left(\det [H_S(k)]\right) .\label{eqn:specwinding}
\end{align}
The genuinely non-Hermitian integer topological invariant $\nu$ characterizes the bulk band-structure of the NTOS by measuring how often the complex phase of the determinant of $H_S(k)$ winds around the origin of the complex plane. Specifically, in Eqs.~(\ref{eqn:one}),(\ref{eqn:deltaE}), we obtain ${\rm Re}[\alpha]>0$, i.e. the desired exponential amplification of the sensitivity $\mathcal S$, if $\lvert \nu\rvert =1$, and ${\rm Re}[\alpha]<0$ implying exponential damping of $\mathcal S$ if $\nu=0$. We note that winding numbers around energies shifted from the origin of the complex plane may readily be defined \cite{gong}. For the particularly experimentally relevant case of a constant imaginary shift to all energies corresponding to a passive setting with loss only \cite{Szameit2017}, this redefinition does not affect the spatial localization and existence of boundary modes.

\begin{figure}[t]
    \centering
    \includegraphics[width=\linewidth]{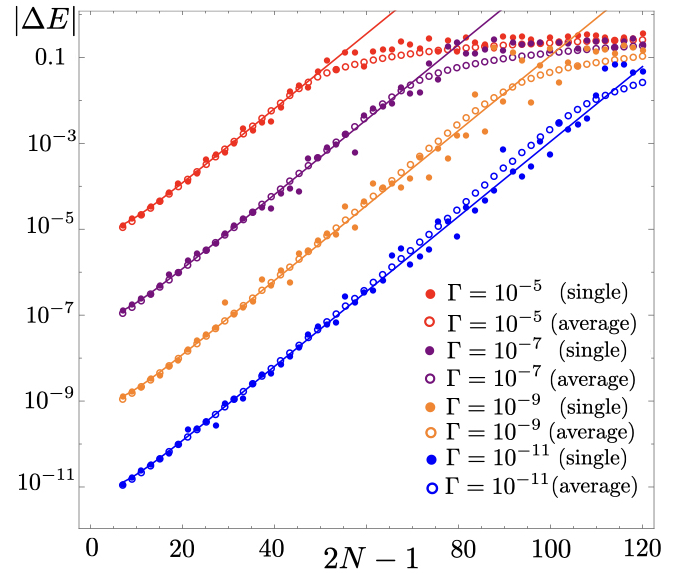}
    \caption{Scaling of the boundary-condition induced energy shift $\lvert \Delta E\rvert$ with system size $2N-1$ in the presence of complex disorder, for various values of $\Gamma$ and system sizes up to $119$ sites. Data for a single disorder realization (solid dots) are compared to a disorder average (circles) over $M=10^4$ disorder realizations at every $N$ \cite{footDisorder}. The disorder strength is $w=0.3$ in both plots. Solid lines indicate the perturbative result (see Eq.~(\ref{eqn:deltaE})) in the absence of disorder.}
    \label{fig2}
\end{figure}

While the qualitative change in the response corresponds to the bulk topological invariant $\nu$ --- and thus occurs at parameters where the Bloch bands touch --- it does not entail gap closings in the open boundary system. In fact, from the perspective of energy gaps there is a well-documented breakdown of the conventional bulk-boundary correspondence: in the open boundary system, gap closings occur at $|t_1^2-\gamma^2|=|t_2|^2$ \cite{BBC,yaowang,KochBBC}, instead of the above condition for the change of $\nu$. The origin of this disparity can be traced back to the zero-modes in Eq.~(\ref{endstates}): the response phase transition (sign change of ${\rm Re}[\alpha]$) occurs when the right {\it or} left eigenstates change their localization, i.e. when $|r_R|=1$ or $|r_L|=1$, whereas the spectral phase transitions are tied to changes in the bi-orthogonal localization occurring at $|r_Lr_R|=1$ \cite{BBC}.  

{\it Stability against local perturbations.---} We now demonstrate the robustness of our main findings against local perturbations. To this end, we add (complex) Gaussian on-site disorder to our above SSH model Hamiltonian of the NTOS. The disorder strength  is quantified by $w>0$. In a wide parameter range, in particular as long as $w$ is significantly smaller than the absolute value of bulk energy gap, we find that such local random perturbations do not have any qualitative effect on the exponential amplification of the sensitivity $\mathcal S$ with system size. Notably, even though the considered disorder breaks the chiral symmetry of the above NH SSH model resulting in a shift of the lowest (in absolute value) energy eigenvalue away from zero, the difference in energy $\Delta E = E_0(\Gamma)-E_0(0)$ induced by a change in boundary conditions is still found to exhibit a similar exponential sensitivity as in the absence of disorder. In Fig.~\ref{fig2}, we corroborate and exemplify this result: Even at the level of individual disorder realizations, the exponential growth of the sensitivity $\mathcal S$ is very clear, unsurprisingly with visible fluctuations. However, we emphasize that these fluctuations are irrelevant for the operation principle of a given NTOS device for which the system size $N$ as well the disorder realization are fixed. In this sense, our data shows that the expected value for the sensitivity of an individual device follows the same rules as without disorder, which becomes clear when averaging over sufficiently many disorder realizations for every system size $N$.

{\it Platforms for implementation.---}
Making available key ingredient for the proposed NTOS, the general physics of anomalous boundary modes in NH topological systems has recently been observed in a range of experimental settings, including mechanical metamaterials \cite{Ghatak}, electrical circuits \cite{ElectricBBC,ElectricBBC2}, and photonic quantum walks \cite{QdynBBC}. For mechanical metamaterials, a detailed dynamical response theory has been developed and shown to posses sharp transitions in the response to external excitations \cite{Henning}. Furthermore, coupling the boundaries of NH systems has been shown to lead to anomalous dynamics \cite{Longhi}. Motivated by these exciting developments, we now outline several scenarios for the implementation of NTOS.

Our main focus is on an array of coupled optical ring-resonators \cite{RingResonatorsBook}, as inspired by experiments on NH EP sensors in such systems \cite{EPsensorExp} as well as optical implementations of topological materials \cite{Hafezi2013,Ozawa2019}. Combining existing technology \cite{Peng2014,Peng2016,EPsensorExp,ChiralQO}, the NH SSH model (\ref{eqn:nhssh}) for our NTOS may be realized as follows \cite{SOM}: The sublattice degree of freedom is represented by the clockwise/anti-clockwise mode of each ring-resonator. The asymmetric coupling between these modes, parameterized by $\gamma$ in Eq.~(\ref{eqn:nhssh}), has been experimentally achieved in Refs.~\cite{Peng2016,EPsensorExp} by introducing two scatterers into the mode-volume of a ring-resonator. While realizing an EP sensor requires fine-tuning of $\gamma$ to compensate the symmetric coupling strength $t_1$, e.g. by adjusting the angle that the scatterers form with the center of the resonator \cite{EPsensorExp}, we stress that the proposed NTOS works in a broad parameter range of finite asymmetry $\gamma$ \cite{SOM}, thus highlighting the topological robustness of our approach. The final ingredient, namely the chirality-selective coupling between neighboring ring-resonators ($t_2$ in Eq.~(\ref{eqn:nhssh})), may be achieved by chiral couplers (see Ref.~\cite{ChiralQO} for a review). Again, far from perfect chirality of the coupling is acceptable to enter the right topological phase of the NTOS, characterized by a non-vanishing winding number $\nu$ (cf. Eq.~(\ref{eqn:specwinding})) \cite{SOM}. In this scenario, the NTOS may be seen as an extension of an EP sensor, alleviating the requirement of fine-tuning and enhancing the attainable precision exponentially in the number coupled ring-resonators (cf. Eq.~(\ref{eqn:one})). Regarding power-consumption and response times, the microscopic similarity of these two devices, in both cases limited by a comparable overall energy scale of $t_1, \gamma$ ($\sim 10$ MHz in Ref.~\cite{EPsensorExp}), allows for their direct comparison. In particular this implies their operation at similar time-scales and similar dissipated energy-per-resonator, respectively.

Beyond this optical realization, in various classical meta-materials, NTOS devices may be constructed from the available toolbox. The probably simplest platform are topological electrical circuits \cite{ElectricBBC,ElectricBBC2}, where basically any topological insulator model in arbitrary geometry may be realized, since arbitrary connectivity of the circuit can readily be achieved. The weak link $\Gamma$ can be designed at any point of a circuit in ring geometry, and a measurand which either couples to an inductance or a capacitor of the device may be detected. By downsizing this architecture to the nano-scale, it may complement the existing toolbox of microelectronic devices including MOSFET sensors. 

The proposed NTOS may also be realized in synthetic quantum materials, such as ultracold atoms in optical lattices \cite{Bloch2008,Goldman2016,Cooper2019} or coupled arrays of quantum dots \cite{loss,Crooker2002,Kagan2016}. In this quantum many-body context, NH Hamiltonians may emerge as an effective description of an underlying quantum master equation \cite{Lindblad1976,BreuerPetruccione}. In particular, the spectral sensitivity of NH matrices has recently been shown \cite{Song2019} to carry over to a full quantum master-equation description corresponding to NH models similar to the above NH SSH model (see Eq.~(\ref{eqn:nhssh})). These insights pave the way towards implementing NTOS in chains of coupled atoms or quantum dots, where the weak link $\Gamma$ coupling to a measurand is simply provided by a high tunnel barrier between two neighboring sites. For the example of quantum dot arrays, such a tunnel barrier may consist of an electrostatic potential, which renders charge measurements a natural application of this quantum version of the proposed NTOS.

{\it Discussion.---} Harnessing the susceptibility of NH topological phases to changes in the boundary conditions, we have identified a robust and widely applicable mechanism for high-precision sensors. In particular, the proposed NTOS devices, designed in an open ring geometry (cf. Fig.~\ref{fig1}), are capable of measuring any observable affecting the coupling between their ends with a sensitivity that increases exponentially with the number of degrees of freedom of the sensor. We have demonstrated this central result by a simple perturbative analysis that is accurately confirmed by the numerically exact solution of a microscopic model, where our findings are largely insensitive to random perturbations in the form of disorder. Furthermore, several platforms for realizing NTOS systems with state of the art experimental methods have been outlined.

On a related note, in a number of recent studies on NH systems, the possibility and implementation of sensors operating at exceptional points, i.e. parameter points where a NH matrix becomes non-diagonalizable, has been extensively discussed 
\cite{EPsensorTheory,EPsensorExp,EPsensorExp2}. There, the energy splitting around an EP of order $m$ at energy $E_e$ scales as $\Delta E \sim \epsilon^{1/m}$, with $\epsilon$ denoting the deviation from the EP, thus also enhancing the sensitivity with respect to $\epsilon$ exponentially strongly in $m$. However, at least two {\textit{crucial differences}} of the EP scenario compared to our present proposal are worth mentioning: First, to prepare an EP of order $m$, an extensive number of $(2 m - 2)$ real parameters \cite{HeissHigher,Hoeller2018} need to be fine-tuned. By contrast, the NTOS systems studied in our present work are largely insensitive to changes in system parameters, except the coupling $\Gamma$ between the boundaries, which is designed to be sensitive to the measurand. Second, the spectral sensitivity of EP sensors has been criticized for not being a suitable quantity to assess the precision of an EP sensor in a noise-limited measurement \cite{Langbein2018, Wang2020}. In particular, the simultaneous $\epsilon$-dependence of the coalescing eigenstates and $\Delta E$ may conspire to yield a linear field response in $\epsilon$, despite the sub-linear $\Delta E$ \cite{Langbein2018}. By contrast, the sensitivity of the proposed NTOS is not stymied by such an interplay between eigenstates and eigenvalues. This is quite intuitive since our scheme relies on the frequency-shift of an energetically isolated mode rather than the coalescence of two modes at an EP.  This intuition is supported by a microscopic analysis of the NH SSH model \cite{SOM} along the lines of \cite{Langbein2018}, clearly showing that the field response of the NTOS system exhibits a similar exponential enhancement with system size as the frequency-shift itself. In this sense, the spectral sensitivity of the NTOS is a suitable quantity to assess the signal to noise ratio that can be experimentally achieved in principle. This point in principle being made, we note that due to the applicability of the NTOS mechanism to a wide range of physical systems, a detailed analysis of the relevant sources of noise for each individual implementation is beyond the scope of this first study, and thus remains a subject of more specialized future work.

The spectral sensitivity at the heart of the NTOS is also intriguing from a viewpoint of boundary modes in topological materials. Normally, in Hermitian systems, hybridization of the edge modes localized at the ends of a topological insulator leads to a mini-gap that {\textit{decays}} exponentially with system size. In the considered NH SSH model instead, coupling the ends of the system leads to a shift in the energy of the boundary mode that {\textit{increases}} exponentially in system size in the parameter regime where the complex bulk energy spectrum of the NTOS winds around the origin of the complex plane. In this sense, non-Hermitian topological sensors are based on a reversed mini-gap mechanism.\\

\acknowledgments
{\it Acknowledgments.---}
We would like to thank Flore Kunst, Alexander Palatnik, Hannes Pichler, and Carsten Timm for discussions. J.C.B. acknowledges financial support from the German Research Foundation (DFG) through the Collaborative Research Centre SFB 1143 and the Cluster of Excellence ct.qmat. E.J.B. is supported by the Swedish Research Council (VR) and the Wallenberg Academy Fellows program of the Knut and Alice Wallenberg Foundation.

\newpage
\begingroup
\onecolumngrid
\appendix
\section{Supplementary Material for ``Non-Hermitian Topological Sensors''}

In this supplementary material, we provide details on our derivations and arguments presented in the main text. First, the analysis of boundary states in NH lattice models and its application to the NH SSH model discussed in the main text is explicated. Second, we analyze a model with on-site dissipation only (i.e. without asymmetric hopping terms) which is found to nevertheless possess an exponential boundary sensitivity, as mentioned in the main text. Third, we detail the implementation of the proposed NTOS devices with coupled optical ring-resonators. Finally, we calculate the response of an NTOS setting, thus confirming our main finding that the exponential amplification of the spectral sensitivity carries over to the full dynamical response.

\subsection{Boundary states in NH lattice models}
We begin by considering generic nearest neighbor hopping models, in particular allowing for asymmetric hopping amplitudes. This includes the non-Hermitian SSH chain studied in the main text (see Fig. \ref{figsub1}(a) for an illustration) as a special case, which is detailed at the end of this section. For periodic boundary conditions, in reciprocal space, the considered models are described by the Bloch Hamiltonian
\begin{align}
H(k) = {\bf d}(k) \cdot \boldsymbol\sigma \qquad {\rm with} \ \ \ \ d_x (k) \pm i d_y (k) \equiv f_\pm + g_\pm {\rm e}^{\mp i k},\ \ \   {\rm where} \ \ f_\pm, g_\pm\in \mathds{C}\ , \label{eqgenblochham}
\end{align}
corresponding to a generic nearest neighbour hopping model with chiral symmetry $\sigma_z H(k) \sigma_z = - H(k)$ as long as $d_z(k)=0$. 
Let us now consider the sensor setup with $2N-1$ sites, for which the Hamiltonian matrix, expanded in the local site basis, is
\begin{equation}
\mathcal{H}= \begin{pmatrix}
0 & f_+ & 0 & 0&  0 &\cdots & \Gamma \\
f_- & 0 & g_+ & 0 & 0 & \cdots &  0 \\
0 & g_-  & 0 &f_+ & 0& \cdots & 0 \\
0 & 0 & f_- &0&  \ddots & \ddots & 0\\
 0 &  0 & 0 &\ddots & \ddots &f_+ & 0\\
 \vdots &  \vdots & \vdots & \ddots & f_- &0 & g_+\\
\Gamma & 0 & 0 & \cdots &0 & g_-  & 0
\end{pmatrix},  \label{eqgeneralhamonerow}
\end{equation}
where $\Gamma$ denotes the hopping between the end sites induced by the measurand. At $\Gamma=0$ there is an exact zero energy ($E=0$) eigenstate of the form 

\begin{equation}
\ket{\Psi_{R,0}}:= \mathcal N_R \begin{pmatrix}
1\\ 0\\ r_R\\ 0\\r_R^2\\0\\r_R^3\\\vdots\\0\\r_R^{N-1}
\end{pmatrix} ; \ \ \  \bra{\Psi_{L,0}}:= \mathcal N_L \begin{pmatrix}
1& 0& r_L& 0&r_L^2&0&r_L^3&\cdots &0& r_L^{N-1}
\end{pmatrix} ,\label{eigentstates}
\end{equation}
with $r_R=-\frac{f_-}{g_+}$ and $r_L=-\frac{f_+}{g_-}$. The overlap between the left and right states is \begin{equation}\braket{\Psi_{L,0}|\Psi_{R,0}}=\mathcal N_L\mathcal N_R\frac{(r_Lr_R)^N - 1}{r_Lr_R-1}\ .\end{equation}
Using this result it is straightforward to calculate the shift in energy induced by the measurand via a coupling of the end sites according to $\Delta H=\Gamma(\ket{1}\bra{2N\!-1}+\ket{2N\!-1}\bra{1})$. To first order perturbation theory in $\Gamma$, we obtain
\begin{align}
\Delta E & \approx \frac{ \bra{\Psi_{L,0}}\Delta H\ket{\Psi_{R,0}}}{\braket{\Psi_{L,0}|\Psi_{R,0}}} = 
\Gamma \frac{(r_L r_R - 1)(r_L^{N-1}+r_R^{N-1})}{(r_Lr_R)^N - 1}
\,~\overset{N \gg 1}{\longrightarrow}~\, \Gamma\, \kappa\,  e^{\alpha N} \ .
\label{eqn:deltaE}
\end{align}
This result, as shown in the main text, provides a very accurate approximation of the energy shift as long as $\Delta E$ is smaller than the energy gap separating the perturbed zero mode from the rest of the spectrum.
Specifically, for the large $N$ limit we find that $\kappa \ne 0$ throughout the considered parameter space,  and we identify 
\begin{equation}
\alpha=\log(\max({r_R, r_L})) \ {\rm when}\ \ \log(|r_Rr_L|))<0 , \ {\rm and} \ \ \alpha=-\log(\min({r_R, r_L})) \ {\rm when}\ \ \log(|r_Rr_L|))>0 \ ,\label{Eshiftdetalis}
\end{equation}
where $\max({r_R, r_L})$ ($\min({r_R, r_L})$) selects the maximum (minimum) value as sorted by the absolute value. Crucially, one finds 
\begin{equation}
{\rm Re}[\alpha]>0 \ {\rm when}\ \ {\rm sign}(\log(|r_R|))\neq {\rm sign}(\log(|r_L|)) , \ {\rm and} \ \ {\rm Re}[\alpha]<0 \ {\rm when}\ \ {\rm sign}(\log(|r_R|))= {\rm sign}(\log(|r_L|)) \label{Eshiftdetalis2}
\end{equation}
implying exponential amplification precisely in the parameter region characterized by a skin effect, i.e. where the perturbed zero mode has right and left eigenvectors located at different ends of the chain.

We end this section by applying our general analysis to the non-Hermitian SSH model with the asymmetric hopping terms (see Fig. \ref{figsub1}(a)), used in our case study in the main text. In reciprocal space, the model with periodic boundaries is described by the Bloch Hamiltonian
\begin{align}
H_S(k) =  \left(t_1 + t_2 \, \textrm{cos} k, \, t_2 \, \textrm{sin}k + i \gamma, \, 0 \right) \cdot \boldsymbol \sigma  \ \  \  {\rm where} \ \ \ t_1, t_2, \gamma\in \mathds{R}\ \ .
\label{eqn:nhsshexplicit}
\end{align}
In the sensor setup with $2N-1$ sites the Hamiltonian matrix is
\begin{equation}
\mathcal{H}= \begin{pmatrix}
0 & t_1-\gamma & 0 & 0&  0 &\cdots & \Gamma \\
t_1+\gamma & 0 & t_2 & 0 & 0 & \cdots &  0 \\
0 & t_2  & 0 &t_1-\gamma & 0& \cdots & 0 \\
0 & 0 & t_1+\gamma &0&  \ddots & \ddots & 0\\
 0 &  0 & 0 &\ddots & \ddots &t_1-\gamma & 0\\
 \vdots &  \vdots & \vdots & \ddots & t_1+\gamma &0 & t_2\\
\Gamma & 0 & 0 & \cdots &0 & t_2  & 0
\end{pmatrix},  \label{eqgeneralhamonerow1}
\end{equation}
and at $\Gamma=0$ there is an exact $E=0$ boundary state of the form of Eq. (\ref{eigentstates}) with $r_R=-\frac{t_1-\gamma}{t_2}$ and $r_L=-\frac{t_1+\gamma}{t_2}$. For this case, the region of exponential signal amplification becomes $||t_1|-|t_2||<|\gamma |<||t_1|+|t_2||$ as quoted in the main text, and which follows from Eq. (\ref{Eshiftdetalis2}).

\begin{figure*} \label{figsub1}
	\centerline{\includegraphics[width=\linewidth]{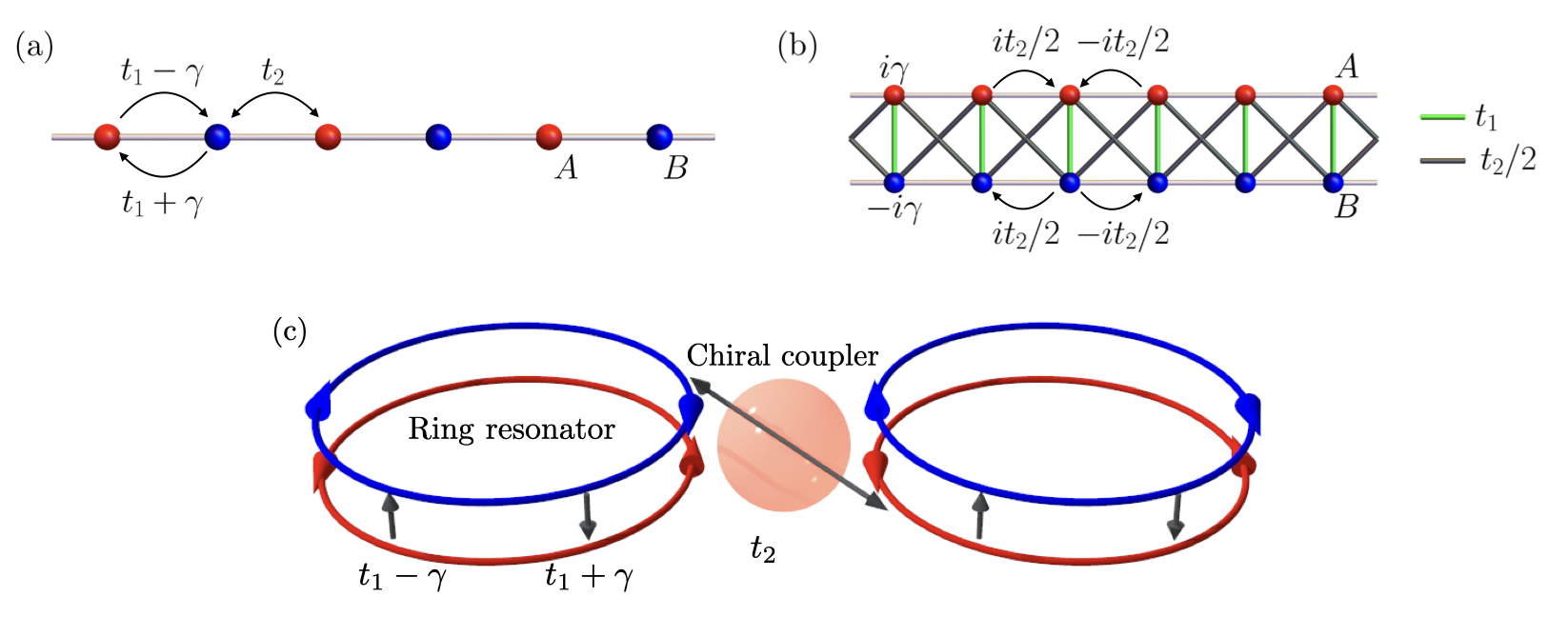}}
	\caption{NTOS models (a) shows the asymmetric hopping model discussed in the main text and (see Eq.~(\ref{eqgeneralhamonerow1})) (b) shows a model whose non-Hermiticity is reflected in an alternating gain and loss (see Eq.~(\ref{eqgeneralhamonerow2})). For an even number of sites the two models are unitary equivalent, while the sensors based on lattices with an odd number of sites behave identically for large enough systems as discussed in text and demonstrated in Fig. \ref{figsub2}. (c) Illustration of the implementation of the couplings $t1,t2,\gamma$ of the NH SSH model shown in (a) with optical ring resonators. Chiral couplers \cite{ChiralQOS} (here shown as a spherical shape pierced by a double-arrow indicating the coupled modes) mediate the direction-selective coupling $t_2$ between neighboring resonators, while the asymmetry $\gamma$ in the intra-resonator coupling of opposite chiralities is experimentally achieved by two scatterers (per ring-resonator) entering the mode volume of the resonators \cite{Peng2016S}. }
	\label{figsub1}
\end{figure*}

\subsection{Sensor with on-site non-Hermitian terms only}

An alternative lattice model platform for the NTOS exclusively featuring diagonal NH terms is shown in Fig. \ref{figsub1}(b). With periodic boundary conditions the Bloch Hamiltonian is given by
\begin{equation}
\tilde{H}(k) = \left(t_1 + t_2 \cos k , \,0,\, t_2 \sin k + i \gamma\right) \cdot \boldsymbol\sigma \ .
\end{equation}
While $\tilde{H}$ has its non-Hermiticity reflected only in a staggered on-site gain and loss pattern, it is unitarily equivalent to the NH SSH model with asymmetric hopping studied in the main text by virtue of the transformation  
\begin{equation}
	\tilde{H} = U^{\dagger}H_SU \ \ {\rm with}\ \ 	U = \frac{1}{\sqrt{2}} \begin{pmatrix} 1 & i \\ i &1 \end{pmatrix}
\end{equation}
and these two models thus have identical spectra for periodic boundary conditions. The unitary equivalence also holds for open boundaries for an even number of sites \cite{Elisabet2020S}, but is not identically satisfied for an odd number of sites. There, the Hamiltonian matrix describing the sensor setup in real space reads as \begin{equation}
\tilde{\mathcal{H}}= \begin{pmatrix}
-i\gamma & t_1 & it_2/2 & t_2/2&  0 &\cdots & \Gamma \\
t_1 & i\gamma & t_2/2 & -it_2/2 & 0 & \cdots &  0 \\
-it_2/2  & t_2/2  & -i\gamma &t_1 & it_2/2 & \cdots & 0 \\
 t_2/2 & it_2/2  & t_1 &i\gamma&  \ddots & \ddots & 0\\
 0 &  0 & -it_2/2 &\ddots & \ddots &t_1 &  it_2/2\\
 \vdots &  \vdots & \vdots & \ddots & t_1 &i\gamma & t_2/2\\
\Gamma & 0 & 0 & \cdots &-i t_2/2 & t_2/2  & -i\gamma
\end{pmatrix},  \label{eqgeneralhamonerow2}
\end{equation}
and does not have a simple exact form of the boundary modes, which for finite systems slightly (exponentially little in system size) differ from zero energy (mini-gap). However, the results from the sensitivity analysis of the NH SSH model still carries over to this system, even to high quantitative accuracy,  once the considered systems are not too small. This finding is illustrated in Fig. \ref{figsub2}, where the analytical results (solid lines) derived for the NH SSH model with asymmetric hopping become essentially identical to the numerical results (dots) for the alternative model ($\tilde H$) with staggered gain and loss.

\begin{figure*} 
	\centerline{\includegraphics[width=0.8\linewidth]{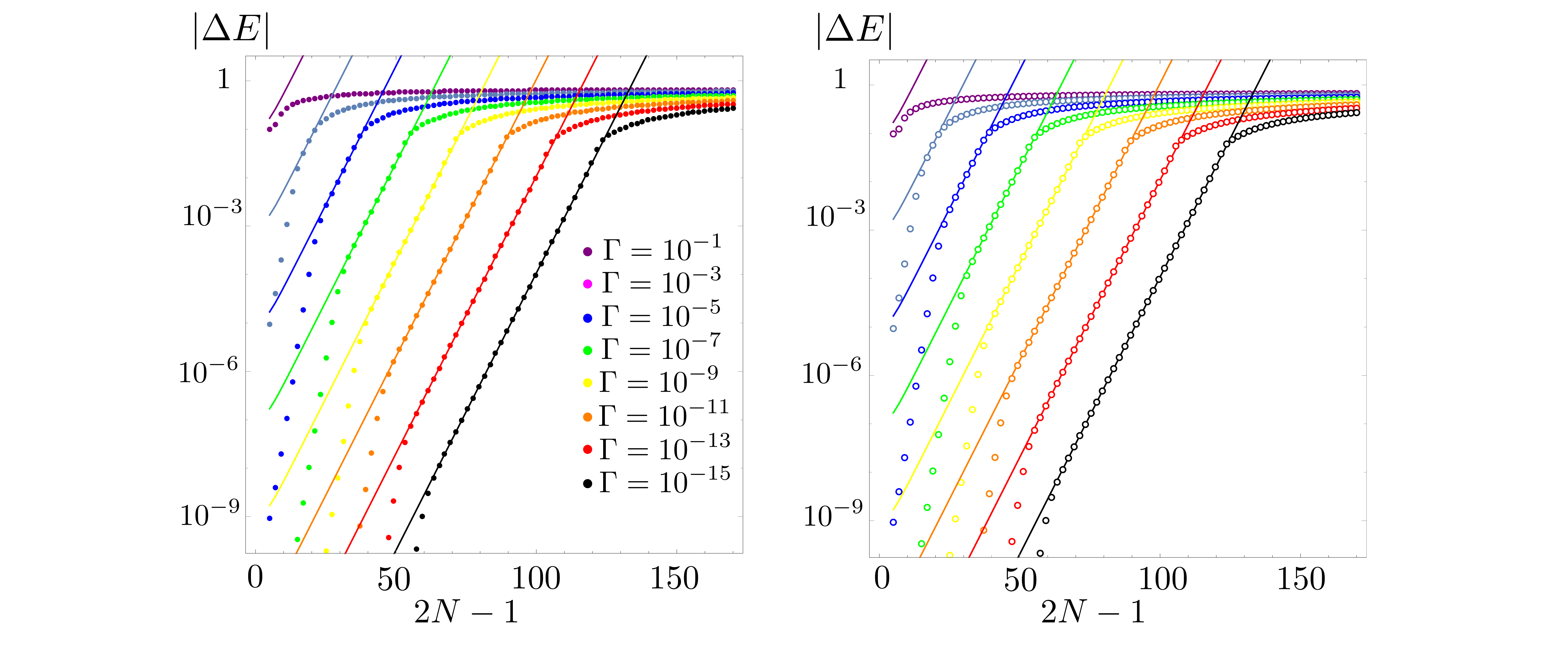}}
	\caption{Exponential dependence of the boundary mode for the model with onsite gain and loss as defined in Eq. \ref{eqgeneralhamonerow2} and illustrated in Fig. \ref{figsub1}(b). The parameters in this example are set to $t_1=t_2=1$ and $\gamma=0.7$. The solid lines represent the perturbative analytical result for the asymmetric hopping model---which, as demonstrated here, becomes essentially exact also for the model with onsite dissipation when the system size is not too small.}\label{figsub2}
\end{figure*}

\subsection{Implementation of NH SSH model with optical ring resonators}
Elaborating on the discussion in the main text, we now describe how the NH SSH model with asymmetric couplings representing the main microscopic model discussed in the main text (cf. Fig.~\ref{figsub2}(a)) can be realized with state of the art experimental techniques. The key ingredients of the proposed implementation are illustrated in Fig.~\ref{figsub2}(c) for a segment of two resonators (unit cells of the NH SSH model). The sublattice degree of freedom of the NH SSH is represented by the two chiralities, clockwise (CL) and counter-clockwise (CO) of the ring-resonator. In the absence of scatterers entering the mode-volume of the resonators, the CL and CO modes may for practical purposes be considered as perfectly decoupled and isolated \cite{Hafezi2013S}. However, by introducing two scatterers into the mode-volume, an asymmetric coupling ($\gamma \ne 0$) has been experimentally achieved \cite{Peng2016S}, where the strength of the asymmetry depends on the difference in the angle between the two scatterers as seen from the center of the ring-resonator. This technology is well under control and even fine-tuning to an exceptional point ($t_1=\gamma$ for a given isolated resonator) has been achieved to high accuracy \cite{Peng2016S,EPsensorExpS}. The core non-Hermitian ingredient $\gamma$ of the proposed NTOS may thus be realized by complete analogy to (multiple copies of) the previously realized EP sensor \cite{EPsensorExpS}, however without any need for fine-tuning (see discussion of phase diagram below). The energy scale of this asymmetric coupling ($\sim 10$ MHz \cite{Peng2016S,EPsensorExpS}) is the limiting factor for the overall time- and energy-scales of both the EP sensor and an NTOS realized on this platform. The (Hermitian) coupling $t_2$ between neighboring resonators is achieved by chiral couplers, a common element from the toolbox of chiral quantum optics \cite{ChiralQOS}. Also here, no fine tuning is required, i.e. the nearest neighbor coupling by no means needs to be perfectly chirality-selective to enter the topologically non-trivial NTOS phase, as quantified below in the discussion of the phase diagram (cf. Fig.~\ref{FigPhaseSup}). These benign aspects of the implementation highlight the topological robustness of our present approach to NH sensing.

Regarding the practical measurement scheme that may be realized with a ring-resonator based NTOS, we note that a straight-forward generalization of the experimental setting of the aforemementioned EP sensors \cite{EPsensorExpS}, amounting to observing a scatterer that perturbs a weak coupling $\Gamma \ll t_2$ realized between two of the ring-resonators (interpreted as the ends of a broken ring geometry), is only one of many conceivable applications of our NTOS. Generally, the weak link $\Gamma$ between the end sites may be represented by any optically impenetrable region imposing a tunneling barrier, i.e. a region with a finite frequency gap hosting only evanescent modes. The measurand then only needs to weakly affect this tunnel barrier, e.g. by slightly changing the refractive index of the impenetrable medium. 

\begin{figure*} 
	\centerline{\includegraphics[width=0.4\linewidth]{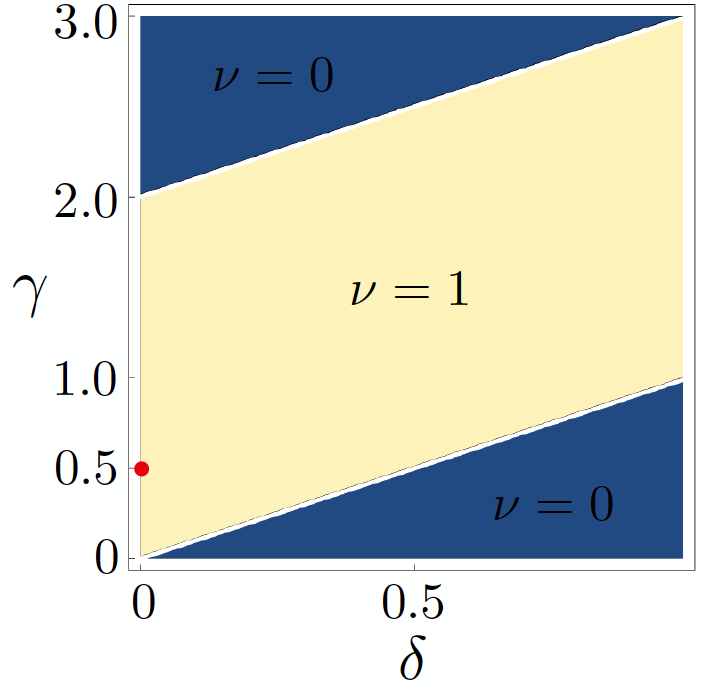}}
	\caption{Phase diagram for the optical ring-resonator implementation of the NH SSH model (see Eq.~(\ref{eqn:imperfectSSH})) as a function of the hopping asymmetry $\gamma$ and the imperfection of the chiral coupler $\delta$. The parameter point discussed in the main text is marked with a red dot. The required phase for operating the NTOS, characterized by the finite winding spectral number $\nu=1$ is shown in light color. The Hermitian hoppings are chosen as $t_1=t_2=1$.}\label{FigPhaseSup}
\end{figure*}

\subsection{Robustness regarding imperfections in the implementation}
To assess and quantify the topological robustness of the NTOS with respect to an imperfect experimental realization of the NH SSH model with the aforementioned ring-resonators (cf. Figs.~\ref{figsub2}(a),(c)), we compute the topological phase diagram (as characterized by the spectral winding number $\nu$). Specifically, we vary the strength of the asymmetry from the value $\gamma=t_1/2 = 0.5$ chosen for our simulations in the main text, and consider an imperfection $\delta$ that quantifies the breaking of the chiral selectivity of the nearest neighbor coupling $t_2$. Regarding the latter, the coupling $\delta$ still respect the chiral symmetry of the NH SSH model that for open boundary conditions ensures zero energy $E_0=0$ of the end state, but couples the CL in a given resonator to the CO mode in its right neighbor, thus amounting to a third-nearest neighbor coupling in Fig.~\ref{figsub2}(a). Overall, in reciprocal space the extended NH Bloch Hamiltonian of the perturbed model reads as
\begin{align}
H(k) =   \left(t_1 + (t_2 + \delta)\, \textrm{cos} k, \, (t_2 - \delta) \, \textrm{sin}k + i \gamma, \, 0\right) \cdot \boldsymbol \sigma.
\label{eqn:imperfectSSH}
\end{align} 
The phase diagram characterized by the value of the spectral winding number $\nu$ (see Eq.~(6) in the main text) is shown in Fig.~\ref{FigPhaseSup}. Clearly, the topologically non-trivial phase ($\nu = 1$) required for the operation of the NTOS extends over a wide parameter regime, basically only requiring that the (at least in the experimentally most relevant regime $\gamma \le t_1$) desirable hopping asymmetry $\gamma$ exceeds the undesirable imperfection of the chiral coupler $\delta$.  

In principle, there may also be connections between modes of the same chirality, e.g. a coupling $\delta_2$ connecting the CL modes in neighboring resonators among each other as well as $\delta_2'$ connecting the CO modes, both amounting to next-nearest neighbor couplings in the language of Fig.~\ref{figsub2}(a). Such couplings explicitly break the chiral symmetry of the NH SSH model and lead to additional terms $\frac{\delta_2 + \delta_2'}{2} \sigma_0 +\frac{\delta_2 - \delta_2'}{2} \sigma_z$ in the Bloch Hamiltonian (\ref{eqn:imperfectSSH}). We stress that the topological invariant $\nu$ does not rely on chiral symmetry and the NTOS would thus clearly exhibit robustness against such perturbations in a finite parameter range as well (see also the related discussion on complex disorder around Fig.~2 in the main text). However, in the proposed implementation the symmetry breaking couplings $\delta_2, \delta_2'$ can even be avoided by simple directional couplers, as has experimentally been demonstrated to very high precision \cite{Hafezi2013S}, thus limiting the practical experimental relevance of such perturbations for the proposed implementation.

\subsection{Sensor response}
Adapting the analysis from Ref. \onlinecite{Langbein2018S}, we now calculate the response of an NTOS to an external probe, thus confirming that the exponential spectral sensitivity studied in the main text is directly visible in the dynamical response of the system. Consider the state $\ket{S(t)}$ of the system with an initial excitation $\ket{x(t)}$ which implies 
\begin{equation}
\ket{S(t)}= \int_{-\infty}^t e^{-iH(t-t')}\ket{x(t')}dt' \ ,
\end{equation}
where the system is assumed to be unperturbed as $t\rightarrow -\infty$. Decomposing this in the eigenmodes of the system yields
\begin{equation}
\ket{S(t)}= \sum_n\int_{-\infty}^t e^{-iE_n(t-t')} \frac{ \ket{\Psi_{R,n}}\bra{\Psi_{L,n}}}{\braket{\Psi_{R,n}|\Psi_{L,n}}} \ket{x(t')}dt' \ ,
\end{equation}
Since the relevant signal of the NTOS results from the perturbed zero mode of the system, we can project to the low-energy sector by considering only 
\begin{equation}
\ket{S(\Gamma,t)}\approx \int_{-\infty}^t e^{-iE_0(t-t')} \frac{ \ket{\Psi_{R,0}}\bra{\Psi_{L,0}}}{\braket{\Psi_{R,0}|\Psi_{L,0}}}\ket{x(t')}dt' \ ,
\end{equation}
where the we have explicitly indicated the crucial dependence on the boundary coupling $\Gamma$. This approximation becomes exact in the limit of a large gap to all other states. For an instant excitation $\ket{x(t)}=\delta(t)\ket{x_0}$, we have  
\begin{equation}
\ket{S(\Gamma,t)}=\theta(t) e^{-iE_0t} \frac{ \ket{\Psi_{R,0}}\bra{\Psi_{L,0}}}{\braket{\Psi_{R,0}|\Psi_{L,0}}}\ket{x_0} \ .
\end{equation}
which is a state (a vector) describing the excited system. To see the strength of the response of the system we need to relate this to a scalar which can for instance be done by defining 

\begin{equation}
\mathcal S(\Gamma,t)\equiv \braket{x_f|S(\Gamma,t)},
\end{equation}
where we have used $\ket{x_0}^T=\ket{x_f}:=\begin{pmatrix}
1& 1& 1& \cdots & 1 & 1
\end{pmatrix}$ for our simulations. In Fig. \ref{figsub3}, we show that the sensitivity with respect to $\Gamma$ of the response, as quantified by the absolute value of $\Delta \mathcal S/\Delta \Gamma=(\mathcal S(\Gamma,t)-\mathcal S(0,t))/\Gamma\approx \frac{\partial  \mathcal S}{\partial \Gamma}$, indeed exhibits the exponential dependence on system size as advertised and anticipated by the corresponding shift in the energy eigenvalue $E_0$. This implies that the spectral sensitivity studied in the main text as the key signature of an NTOS indeed gives a good picture of the precision that may be achieved in an actual response experiment \cite{Langbein2018S}.   

\begin{figure*} 
	\centerline{\includegraphics[width=0.4\linewidth]{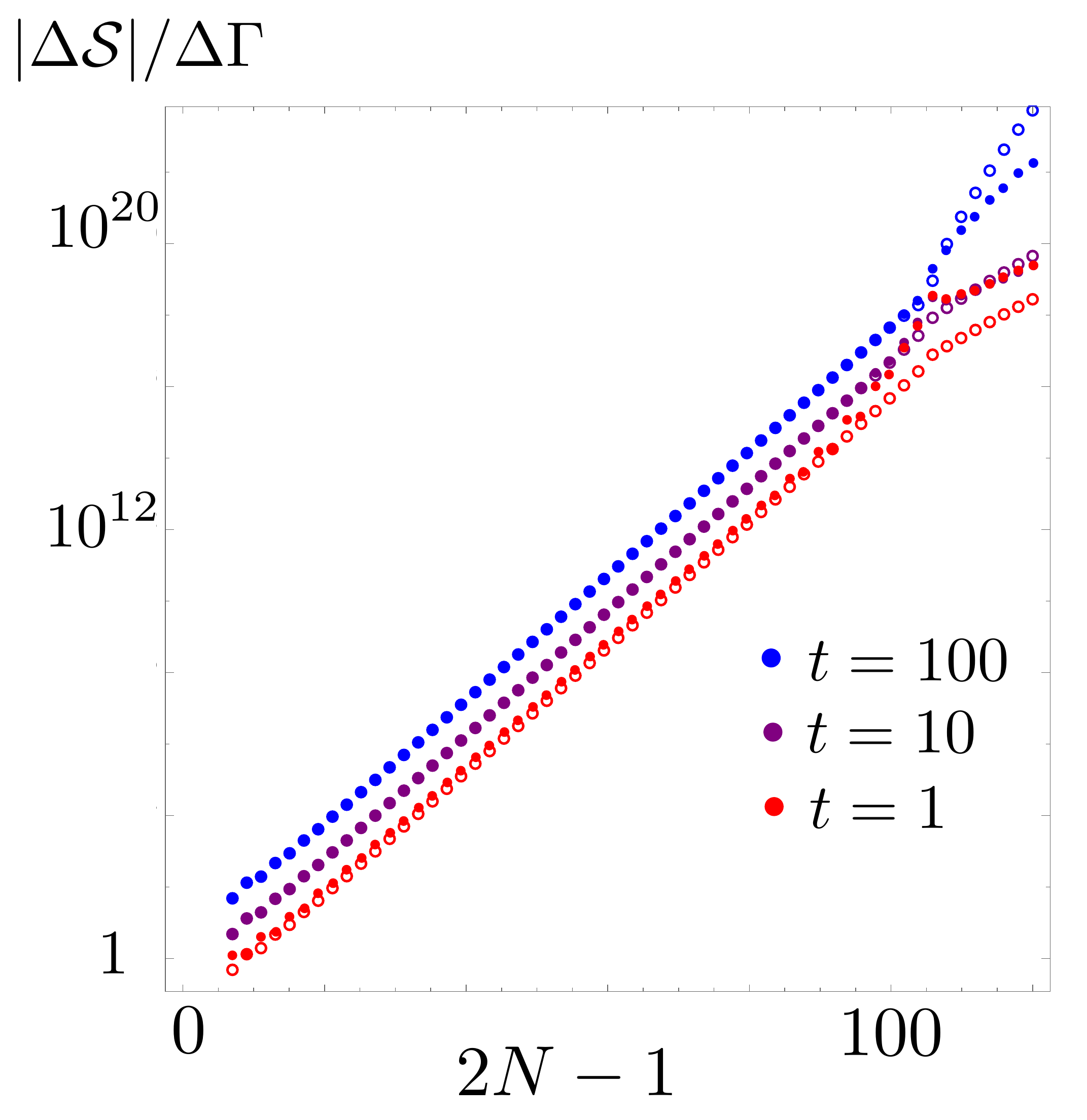}}
	\caption{Sensor performance for the non-Hermitian SSH model defined in Eq. \ref{eqgeneralhamonerow1} which is also the model studied in the main text. The absolute value of $\Delta \mathcal S/\Delta \Gamma=(\mathcal S(\Gamma,t)-\mathcal S(0,t))/\Gamma\approx \frac{\partial  \mathcal S}{\partial \Gamma}$ with a small coupling, here $\Gamma=10^{-10}$, for various times, $t$, exhibits an essentially perfect exponential increase with system size for many orders of magnitude until the system is so big that the energy shift becomes comparable to the gap in the system. A second important feature is that the response accounting for the $\Gamma$-dependence of both eigen-energies and eigen-states (solid dots), is essentially on top of the approximate result (rings, partly covered by the full result), for which only the change in the energy eigenvalues enters (thus ignoring the change in the eigenstates), as considered in the main text. A significant difference is only seen for very large systems and at short times, where the behaviour is nevertheless qualitatively similar. 
	The model parameters are $t_1=t_2=\gamma=1$.}\label{figsub3}
\end{figure*}

\endgroup


\begin{thebibliography}{10}

\bibitem{cavitysensor}K. D. Heylman,  N. Thakkar, E. H. Horak, S. C. Quillin, C. Cherqui, K. A. Knapper, D. J. Masiello, and R. H. Goldsmith, {\em Optical microresonators as single-particle absorption spectrometers}, \href{https://www.nature.com/articles/nphoton.2016.217}{Nat. Photon. {\bf 10}, 788 (2016)}.

\bibitem{cavitysensor2} J. Zhu, S. K. Ozdemir, Y.-F. Xiao, L. Li, L. He, D.-R. Chen and L. Yang, {\em On-chip single nanoparticle detection and sizing by mode splitting in an ultrahigh-Q microresonator}, \href{https://www.nature.com/articles/nphoton.2009.237}{Nat. Photon. {\bf 4}, 46 (2010)}.

\bibitem{cavitysensor3} F. Vollmer and L. Yang, {\em Label-free detection with high-Q microcavities: a review of biosensing mechanisms for integrated devices}, \href{https://www.degruyter.com/view/journals/nanoph/1/3-4/article-p267.xml}{Nanophotonics {\bf 1}, 267 (2012)}.

\bibitem{cavitysensor4} L. He, S.K. \"Ozdemir, J. Zhu, W. Kim and L. Yang , {\em Detecting single viruses and nanoparticles using whispering gallery microlasers}, \href{https://www.nature.com/articles/nnano.2011.99}{Nat. Nanotechnol. {\bf 6}, 428 (2011)}.

\bibitem{singleatomionsensor}M. D. Baaske and F. Vollmer, 
 {\em Optical observation of single atomic ions interacting with plasmonic nanorods in aqueous solution}, \href{https://www.nature.com/articles/nphoton.2016.177}{Nat. Photon. {\bf 10}, 733 (2016)}.

\bibitem{MOSFET} P. Bergveld, {\em The impact of MOSFET-based sensors}, \href{https://doi.org/10.1016/0250-6874(85)87009-8}{Sensors and Actuators {\bf 8}, 109 (1985)}.

\bibitem{mechtrand}  E. Gavartin, P. Verlot, and T.J. Kippenberg, {\em A hybrid on-chip optomechanical transducer for ultrasensitive force measurements}, \href{https://www.nature.com/articles/nnano.2012.97}{Nat. Nanotechnol. {\bf 7}, 509 (2012)}.

\bibitem{graphenesensor}
E. W. Hill, A. Vijayaragahvan and K. Novoselov,  {\em Graphene Sensors}, \href{https://ieeexplore.ieee.org/document/6016205}{IEEE Sensors Journal, {\bf 11}, 3161, (2011)}.

\bibitem{Squids} J. Clarke and A. I. Braginski, eds. {\em The SQUID handbook: Applications of SQUIDs and SQUID systems}, \href{https://onlinelibrary.wiley.com/doi/book/10.1002/3527603646}{John Wiley \& Sons (2006)}.

\bibitem{magnetometers} S. Forstner, S. Prams, J. Knittel, E. D. van Ooijen, J. D. Swaim, G. I. Harris, A. Szorkovszky, W. P. Bowen, and H. Rubinsztein-Dunlop, {\em Cavity optomechanical magnetometer}, \href{https://journals.aps.org/prl/abstract/10.1103/PhysRevLett.108.120801}{Phys. Rev. Lett. {\bf 108}, 120801 (2012)}.

\bibitem{BerryDeg}
M. Berry,  {\em Physics of Nonhermitian Degeneracies}, \href{https://link.springer.com/article/10.1023\%2FB\%3ACJOP.0000044002.05657.04}{Czechoslovak Journal of Physics {\bf 54}, 1039 (2004)}.

\bibitem{Heiss}
W. D. Heiss, {\em The physics of exceptional points}, \href{http://adsabs.harvard.edu/abs/2012JPhA...45R4016H}{Journal of Physics A. {\bf 45}, 444016 (2012)}.

\bibitem{EPreview}
M.-A. Miri and A. Alu,  {\em Exceptional points in optics and photonics}, \href{https://science.sciencemag.org/content/363/6422/eaar7709}{Science {\bf 363}, eaar7709 (2019)}.


\bibitem{BBC}
F.K. Kunst, E. Edvardsson, J.C. Budich, and E.J. Bergholtz, {\em Biorthogonal Bulk-Boundary Correspondence in non-Hermitian Systems}, \href{https://journals.aps.org/prl/abstract/10.1103/PhysRevLett.121.026808}{Phys. Rev. Lett. {\bf 121}, 026808 (2018)}.

\bibitem{lee}
T.E. Lee, {\em Anomalous Edge State in a Non-Hermitian Lattice}, \href{https://journals.aps.org/prl/abstract/10.1103/PhysRevLett.116.133903}{Phys. Rev. Lett. {\bf 116}, 133903 (2016)}.

\bibitem{Xi2018}
Y. Xiong, {\em Why does bulk boundary correspondence fail in some non-Hermitian topological models}, \href{http://iopscience.iop.org/article/10.1088/2399-6528/aab64a/meta}{J. Phys. Commun. {\bf 2}, 035043 (2018)}.

\bibitem{yaowang}
S. Yao and Z. Wang, {\em Edge states and topological invariants of non-Hermitian systems}, \href{https://doi.org/10.1103/PhysRevLett.121.086803}{Phys. Rev. Lett. {\bf 121}, 086803 (2018)}.

\bibitem{Henning}H. Schomerus, {\em Nonreciprocal response theory of non-Hermitian mechanical metamaterials: Response phase transition from the skin effect of zero modes}, \href{https://journals.aps.org/prresearch/abstract/10.1103/PhysRevResearch.2.013058}{Phys. Rev. Research {\bf 2}, 013058 (2020)}.

\bibitem{EPsensorTheory}
J. Wiersig, {\em Enhancing the Sensitivity of Frequency and Energy Splitting Detection by Using Exceptional Points: Application to Microcavity Sensors for Single-Particle Detection}, \href{https://journals.aps.org/prl/abstract/10.1103/PhysRevLett.112.203901}{Phys. Rev. Lett. {\bf 112}, 203901 (2014)}.

\bibitem{EPsensorExp}
H. Hodaei, A. U. Hassan, S. Wittek, H. Garcia-Gracia, R. El-Ganainy, D. N. Christodoulides and M. Khajavikhan, {\em Enhanced sensitivity at higher-order exceptional points}, \href{https://www.nature.com/articles/nature23280}{Nature {\bf 548}, 187 (2017)}.

\bibitem{EPsensorExp2}
W. Chen, S.K. \"Ozdemir, G. Zhao, J. Wiersig, and L. Yang, {\em Exceptional points enhance sensing in an optical microcavity}, \href{https://www.nature.com/articles/nature23281}{Nature {\bf 548}, 192 (2017)}.

\bibitem{Ghatak} A. Ghatak, M. Brandenbourger, J. van Wezel, and C. Coulais, {\em Observation of non-Hermitian topology and its bulk-edge correspondence}, \href{https://arxiv.org/abs/1907.11619}{arXiv:1907.11619}.


\bibitem{ElectricBBC} T. Helbig, T. Hofmann, S. Imhof, M. Abdelghany, T. Kiessling, L. W. Molenkamp, C. H. Lee, A. Szameit, M. Greiter, and R. Thomale, {\em Observation of bulk boundary correspondence breakdown in topolectrical circuits}, \href{https://arxiv.org/abs/1907.11562}{1907.11562}.

\bibitem{ElectricBBC2}
T. Hofmann, T. Helbig, F. Schindler, N. Salgo, M. Brzezinska, M. Greiter, T. Kiessling, D. Wolf, A. Vollhardt, A. Kabasi, C. H. Lee, A. Bilusic, R. Thomale, and T. Neupert, {\em Reciprocal skin effect and its realization in a topolectrical circuit}, \href{https://arxiv.org/abs/1908.02759}{arXiv:1908.02759}.

\bibitem{QdynBBC}
L. Xiao, T. Deng, K. Wang, G. Zhu, Z. Wang, W. Yi, and P. Xue , {\em Non-Hermitian bulk-boundary correspondence in quantum dynamics}, \href{https://www.nature.com/articles/s41567-020-0836-6}{Nat. Phys. (2020)}.


\bibitem{SzameitScience2020}
S. Weidemann, M. Kremer, T. Helbig, T. Hofmann, A. Stegmaier, M. Greiter, R. Thomale, A. Szameit, {\em Topological funneling of light}, \href{https://science.sciencemag.org/content/368/6488/311/tab-article-info}{
Science {\bf{368}}, 311 (2020)}.


\bibitem{review}
 E.J. Bergholtz, J.C. Budich, and F.K. Kunst, {\em Exceptional Topology of Non-Hermitian Systems}, \href{https://arxiv.org/abs/1912.10048}{arxiv:1912.10048}.

\bibitem{gong}
Z. Gong, Y. Ashida, K. Kawabata, K. Takasan, S. Higashikawa, and M. Ueda, {\em Topological Phases of Non-Hermitian Systems}, \href{https://journals.aps.org/prx/abstract/10.1103/PhysRevX.8.031079}{Phys. Rev. X {\bf 8}, 031079 (2018)}.

\bibitem{NHarc}
L. Lu, Z. Wang, D. Ye, L. Ran, L. Fu, J. D. Joannopoulos, and M. Solja\v{c}i\'{c}, {\em Observation of bulk Fermi arc and polarization half charge from paired exceptional points}, \href{http://science.sciencemag.org/content/
359/6379/1009}{Science {\bf 359}, 1009 (2018)}.

\bibitem{Budich2019}
J. C. Budich, J. Carlstr\"om, F. K. Kunst, and E. J. Bergholtz, {\em Symmetry-protected nodal phases in non-Hermitian systems}, \href{https://journals.aps.org/prb/abstract/10.1103/PhysRevB.99.041406}{Phys. Rev. B {\bf{99}}, 041406(R) (2019)}.

\bibitem{Yoshida2019}
T. Yoshida, R. Peters, N. Kawakami, and Y. Hatsugai, {\em Symmetry-protected exceptional rings in two-dimensional correlated systems with chiral symmetry}, \href{https://journals.aps.org/prb/abstract/10.1103/PhysRevB.99.121101}{Phys. Rev. B {\bf{99}}, 121101(R)  (2019)}.

\bibitem{EPringExp}
A. Cerjan, S. Huang, K. P. Chen, Y. Chong, and M. C. Rechtsman, {\em Experimental realization of a Weyl exceptional ring}, \href{https://www.nature.com/articles/s41566-019-0453-z}{Nat. Photon. {\bf 13}, 623 (2019)}.

\bibitem{Kawabata2019}
K. Kawabata, K. Shiozaki, M. Ueda, and M. Sato, {\em Symmetry and Topology in Non-Hermitian Physics},
\href{https://journals.aps.org/prx/abstract/10.1103/PhysRevX.9.041015}{Phys. Rev. X {\bf{9}}, 041015 (2019)}.

\bibitem{footOdd} In this work, we generally consider systems with an odd number of sites that stabilize a {\em single} low-energy bound state. For an even number of sites, in similar systems a pair of bound states emerges, where coalescence as well as hybridization of those (see Ref.~\cite{KochBBC} for a detailed discussion) leads to deviations from the proposed NTOS scenario.

\bibitem{Albert2015}
V. V. Albert, L. I. Glazman, and L. Jiang, {\em Topological properties of linear circuit lattices}, \href{https://journals.aps.org/prl/abstract/10.1103/PhysRevLett.114.173902}{Phys. Rev. Lett. {\bf{114}}, 173902 (2015)}.

\bibitem{loss}D. Loss and D. P. DiVincenzo, {\em Quantum computation with quantum dots},
\href{https://journals.aps.org/pra/abstract/10.1103/PhysRevA.57.120}{Phys. Rev. A {\bf 57}, 120 (1998).}
%
\bibitem{Crooker2002}
S. A. Crooker, J. A. Hollingsworth, S. Tretiak, and V. I. Klimov, {\em Spectrally Resolved Dynamics of Energy Transfer in Quantum-Dot Assemblies: Towards Engineered Energy Flows in Artificial Materials},
\href{https://journals.aps.org/prl/abstract/10.1103/PhysRevLett.89.186802}{Phys. Rev. Lett. {\bf{89}}, 186802 (2002)}.

\bibitem{Kagan2016}
C. R. Kagan, E. Lifshitz, E. H. Sargent, D. V. Talapin, {\em Building devices from colloidal quantum dots}, \href{https://science.sciencemag.org/content/353/6302/aac5523.full}{Science {\bf{353}}, 5523 (2016)}


\bibitem{Bloch2008}
I.~Bloch, J.~Dalibard, W.~Zwerger,
{\em Many-body physics with ultracold gases},
\href{https://journals.aps.org/rmp/abstract/10.1103/RevModPhys.80.885}{Rev. Mod. Phys. {\bf 80}, 885 (2008).}

\bibitem{Goldman2016}
N.~Goldman, J.~C.~Budich, and P.~Zoller, {\em Topological quantum matter with ultracold gases in optical lattices},
\href{https://www.nature.com/articles/nphys3803}{Nat. Phys. {\bf 12}, 639 (2016).}
%
\bibitem{Cooper2019}
N.~R.~Cooper, J.~Dalibard, and I.~B.~Spielman, {\em Topological bands for ultracold atoms},
\href{https://link.aps.org/doi/10.1103/RevModPhys.91.015005}{Rev. Mod. Phys. {\bf 91}, 015005 (2019). }
%
\bibitem{SOM}
See the supplementary online material for technical details and extended derivations.
%
\bibitem{ssh1979}
W. P. Su, J. R. Schrieffer, and A. J. Heeger, {\em Solitons in Polyacetylene}, \href{https://journals.aps.org/prl/abstract/10.1103/PhysRevLett.42.1698}{Phys. Rev. Lett. {\bf{42}}, 1698 (1979)}.
%
\bibitem{Lieu2018}
S. Lieu, {\em Topological phases in the non-Hermitian Su-Schrieffer-Heeger model}, \href{https://journals.aps.org/prb/abstract/10.1103/PhysRevB.97.045106}{Phys. Rev. B {\bf{97}} (4), 45106 (2018)}.
%
\bibitem{footModel}
We choose to illustrate the proposed NTOS by example of the well-studied NH SSH model which exhibits off-diagonal NH terms in the form of non-reciprocal hopping. This is however not a necessary ingredient for the exponentially enhanced sensitivity at the heart of NTOS. In particular, there are microscopic models exhibiting a similar sensitivity of topological bound state energies which exclusively feature diagonal NH terms \cite{SOM}.
%
\bibitem{Szameit2017}
S. Weimann, M. Kremer, Y. Plotnik, Y. Lumer, S. Nolte, K. G. Makris, M. Segev, M. C. Rechtsman, A. Szameit, {\em Topologically protected bound states in photonic parity–time-symmetric crystals}, \href{https://www.nature.com/articles/nmat4811}{Nat. Mater. {\bf{16}}, 433 (2017)}.
%
\bibitem{footDisorder} To account for possible spectral degeneracies, the two lowest modes of the system both at finite $\Gamma$ and with open boundaries are considered, and $\Delta E$ is taken as the difference with the smallest absolute value within this manifold.

\bibitem{KochBBC}
R. Koch  and J. C. Budich, {\em Bulk-boundary correspondence in non-Hermitian systems: stability analysis for generalized boundary conditions}, \href{https://arxiv.org/abs/1912.07687}{arXiv e-prints arXiv:1912.07687}.

\bibitem{Longhi}S. Longhi, {\em Loschmidt Echo and Fidelity Decay Near an Exceptional Point}, \href{https://onlinelibrary.wiley.com/doi/abs/10.1002/andp.201900054}{Ann. Phys. {\bf 531} 1900054 (2019)}.

\bibitem{RingResonatorsBook}
V. Van, {\emph{Optical Microring Resonators}}, CRC Press (2017).

\bibitem{Hafezi2013}
M. Hafezi, S. Mittal, J. Fan, A. Migdall, and J. M. Taylor, Nat. Photonics {\bf{7}}, 1001 (2013).


\bibitem{Ozawa2019}
T. Ozawa, H. M. Price, A. Amo, N. Goldman, M. Hafezi, L. Lu, M. C. Rechtsman, D. Schuster, J. Simon, O. Zilberberg, and I. Carusotto, {\em Topological photonics},
\href{https://journals.aps.org/rmp/abstract/10.1103/RevModPhys.91.015006}{Rev. Mod. Phys. {\bf{91}}, 015006 (2019)}.

\bibitem{Peng2014}
B. Peng, {\c S}. K. {\"O}zdemir, F. Lei, F. Monifi, M. Gianfreda, G. L. Long, S. Fan, F. Nori, C. M. Bender, and L. Yang, {\em Parity–time-symmetric whispering-gallery microcavities}, \href{https://www.nature.com/articles/nphys2927}{Nat. Phys. {\bf{10}}, 394–398 (2014)}.

\bibitem{Peng2016}
B. Peng, S. K. \"Ozdemir, M. Liertzer, W. Chen, J. Kramer, H. Yilmaz, J. Wiersig, S. Rotter, and L. Yang, {\em Chiral modes and directional lasing at exceptional points},\href{https://www.pnas.org/content/113/25/6845}{PNAS {\bf{113}}, 6845 (2019).} 

\bibitem{ChiralQO}
P. Lodahl, S. Mahmoodian, S. Stobbe, A. Rauschenbeutel, P. Schneeweiss, J. Volz, H. Pichler, and P. Zoller, {\em Chiral quantum optics}, \href{https://www.nature.com/articles/nature21037}{Nature {\bf{541}}, 473 (2017).} 

\bibitem{Lindblad1976}
G. Lindblad, {\em On the generators of quantum dynamical semigroups}, \href{https://link.springer.com/article/10.1007/BF01608499}{Commun. Math. Phys. {\bf{48}} (2), 119–130 (1976)}.

\bibitem{BreuerPetruccione}
H.-P. Breuer and F. Petruccione, \href{http://www.oup.com/localecatalogue/google/?i=9780198520634}{\em The theory of open
quantum systems} (Oxford University Press) (2002).

\bibitem{Song2019}
F. Song, S. Yao, and Z. Wang, {\em Non-Hermitian Skin Effect and Chiral Damping in Open Quantum Systems}, \href{https://journals.aps.org/prl/abstract/10.1103/PhysRevLett.123.170401}{Phys. Rev. Lett. {\bf{123}} (17), 170401 (2019)}.

\bibitem{HeissHigher}
W. D. Heiss, {\em Chirality of wavefunctions for three coalescing levels}, \href{https://iopscience.iop.org/article/10.1088/1751-8113/41/24/244010}{Journal of Physics A. {\bf 41}, 244010 (2008)}.

\bibitem{Hoeller2018}
J. H\"oller, N. Read, J. G. E. Harris, {\em Non-Hermitian adiabatic transport in the space of exceptional points}, \href{https://arxiv.org/abs/1809.07175}{arXiv:1809.07175 (2018).}

\bibitem{Langbein2018}
W. Langbein, {\em No exceptional precision of exceptional-point sensors}, \href{https://journals.aps.org/pra/pdf/10.1103/PhysRevA.98.023805}{Phys. Rev. A {\bf{98}}, 023805 (2018)}.

\bibitem{Wang2020}
H. Wang, Y.-H. Lai, Z. Yuan, M.-G. Suh, and K. Vahala, {\em Petermann-factor sensitivity limit near an exceptional point in a Brillouin ring laser gyroscope}, \href{https://www.nature.com/articles/s41467-020-15341-6}{Nat. Commun. {\bf{11}},1610 (2020)}.


\end{thebibliography}

\begin{thebibliography}{100}
\bibitem[S1]{Elisabet2020S}
E. Edvardsson, F. K. Kunst, T. Yoshida, and E.J. Bergholtz, {\em Phase transitions and generalized biorthogonal polarization in non-Hermitian systems}, \href{https://journals.aps.org/prresearch/abstract/10.1103/PhysRevResearch.2.043046}{Phys. Rev. Research {\bf 2}, 043046 (2020)}.

\bibitem[S2]{Hafezi2013S}
M. Hafezi, S. Mittal, J. Fan, A. Migdall, and J. M. Taylor, Nat. Photonics {\bf{7}}, 1001 (2013).
\bibitem[S3]{Peng2016S}
B. Peng, S. K. \"Ozdemir, M. Liertzer, W. Chen, J. Kramer, H. Yilmaz, J. Wiersig, S. Rotter, and L. Yang, {\em Chiral modes and directional lasing at exceptional points},\href{https://www.pnas.org/content/113/25/6845}{PNAS {\bf{113}}, 6845 (2019).} 
\bibitem[S4]{EPsensorExpS}
H. Hodaei, A. U. Hassan, S. Wittek, H. Garcia-Gracia, R. El-Ganainy, D. N. Christodoulides and M. Khajavikhan, {\em Enhanced sensitivity at higher-order exceptional points}, \href{https://www.nature.com/articles/nature23280}{Nature {\bf 548}, 187 (2017)}.
\bibitem[S5]{ChiralQOS}
P. Lodahl, S. Mahmoodian, S. Stobbe, A. Rauschenbeutel, P. Schneeweiss, J. Volz, H. Pichler, and P. Zoller, {\em Chiral quantum optics}, \href{https://www.nature.com/articles/nature21037}{Nature {\bf{541}}, 473 (2017).} 
\bibitem[S6]{Langbein2018S}
W. Langbein, {\em No exceptional precision of exceptional-point sensors}, \href{https://journals.aps.org/pra/pdf/10.1103/PhysRevA.98.023805}{Phys. Rev. A {\bf{98}}, 023805 (2018)}.
\end{thebibliography}
\end{document}